\documentclass[10pt,journal,compsoc]{IEEEtran}

\usepackage{graphicx}
\usepackage{subfigure}
 \usepackage{algorithm}
\usepackage{algorithmic}
\usepackage{booktabs} 
\usepackage{amsmath}
 \usepackage{multirow}
 \usepackage{rotating}
 \usepackage{amssymb}
 \usepackage{bbding}
 \usepackage{amsthm}
 \usepackage{bm}
 \usepackage{epstopdf}
 \usepackage{enumerate}
 \usepackage{array}
\usepackage{hyperref}
\usepackage{epstopdf}

\ifCLASSOPTIONcompsoc
  \usepackage[nocompress]{cite}
\else
  \usepackage{cite}
\fi

\newcommand{\Set}[1]{\mathcal{#1}}

\newcommand{\ie}{\emph{i.e., }}
\newcommand{\eg}{\emph{e.g., }}
\newcommand{\etal}{\emph{et al.}}

\ifCLASSINFOpdf
\else
\fi

\begin{document}

\title{Popularity Bias Is Not Always Evil: Disentangling Benign and Harmful Bias for Recommendation}

\author{Zihao Zhao,
        Jiawei Chen, Sheng Zhou, Xiangnan He, Xuezhi Cao, Fuzheng Zhang, Wei Wu \IEEEcompsocitemizethanks{\IEEEcompsocthanksitem Zihao Zhao,
        Jiawei Chen and Xiangnan He are with the School
of Information Science and Technology, University of Science and Technology of China, Hefei, China.
E-mail: zzh1998@mail.ustc.edu.cn, cjwustc@ustc.edu.cn and hexn@ustc.edu.cn.
}
\IEEEcompsocitemizethanks{\IEEEcompsocthanksitem Sheng Zhou is with the College of Computer Science, Zhejiang University, Hangzhou, China. E-mail: zhousheng\_zju@zju.edu.cn.}
\IEEEcompsocitemizethanks{\IEEEcompsocthanksitem Xuezhi Cao, Fuzheng Zhang, Wei Wu are with Meituan Dianping, Beijing, China. E-mail: caoxuezhi@meituan.com, zhangfuzheng@meituan.com, wuwei30@meituan.com.}
\IEEEcompsocitemizethanks{\IEEEcompsocthanksitem Jiawei Chen is the corresponding author.}

}


\markboth{Transactions on Knowledge and Data Engineering}%
{Shell \MakeLowercase{\textit{et al.}}: Bare Demo of IEEEtran.cls for Computer Society Journals}

\IEEEtitleabstractindextext{%
\begin{abstract}
Recommender system usually suffers from severe \textit{popularity bias} --- the collected interaction data usually exhibits quite imbalanced or even long-tailed distribution over items. Such skewed distribution may result from the users' \textit{conformity} to the group, which deviates from reflecting users' true preference. Existing efforts for tackling this issue mainly focus on completely eliminating popularity bias. However, we argue that not all popularity bias is evil. Popularity bias not only results from conformity but also \textit{item quality}, which is usually ignored by existing methods. Some items exhibit higher popularity as they have intrinsic better property. Blindly removing the popularity bias would lose such important signal, and further deteriorate model performance. To sufficiently exploit such important information for recommendation, it is essential to disentangle the benign popularity bias caused by item quality from the harmful popularity bias caused by conformity.

 Although important, it is quite challenging as we lack an explicit signal to differentiate the two factors of popularity bias. In this paper, we propose to leverage temporal information as the two factors exhibit quite different patterns along the time: item quality revealing item inherent property is stable and static while conformity that depends on items' recent clicks is highly time-sensitive. Correspondingly, we further propose a novel \textbf{Ti}me-aware \textbf{D}is\textbf{E}ntangled framework (\textbf{TIDE}), where a click is generated from three components namely the static item quality, the dynamic conformity effect, as well as the user-item matching score returned by any recommendation model. Lastly, we conduct interventional inference such that the recommendation can benefit from the benign popularity bias while circumvent the harmful one. Extensive experiments on three real-world datasets demonstrated the effectiveness of TIDE.
\end{abstract}

}

\maketitle

\IEEEdisplaynontitleabstractindextext

\IEEEpeerreviewmaketitle

\IEEEraisesectionheading{\section{Introduction}\label{sec:introduction}}
Recent years have witnessed flourishing publications on recommendation, most of which aim at inventing machine learning models to fit users' historical behavior data \cite{wu2021survey}. However, the observation data usually exhibits severe \textit{popularity bias}, \ie the distribution over items is quite imbalanced and even long-tailed. Such skewed distribution may be caused by the users' \textit{conformity}, deviating from reflecting users' true preference. As a crucial factor for user decision-making, conformity describes the tendency that user behaves following groups. In a typical recommender system, a user may click an item simply because he finds the item clicked by many other users, rather than based on his own judgement. As a result, recommendation model trained on such biased data would yield unexpected results, e.g., capturing skewed user preference and amplifying the long-tail effect.

Given the wide existence of popularity bias and its negative impact on recommendation, we cannot emphasize too much the importance of tackling popularity bias. Existing efforts mainly focus on entirely eliminating popularity bias to recover true user preference. However, we argue that not all popularity bias is harmful. Besides conformity effect, the uneven item distribution can also be attributed to the diverse item quality. For example, some items exhibit higher popularity as they have intrinsic better properties, e.g., attractive story, harmonious music and professional actors for a typical movie. Blindly removing the popularity bias would lose such important signal, making the model fail to differentiate superb items that deserve more opportunities for recommendation. Therefore, we arrive at a dilemma: eliminating popularity bias would lose important quality signal, while maintaining popularity bias would suffer undesirable conformity effect. Now a question is raised: \textit{is there a solution that enjoys the merit of the popularity bias while circumvents its bad effect?} To achieve this goal, it is essential to disentangle the harmful popularity bias caused by the conformity from the benign one caused by the item quality.

Although important, this problem has been under explored on the literature. The main challenge is the lack of explicit signals for disentanglement. Since we only have access to item popularity scores, which do not tell what factor causes this result. To deal with this problem, we propose to leverage the temporal information in differentiating the benign and harmful factors, as they exhibit quite different patterns along time: item quality which reveals item intrinsic property is stable and static, while conformity that depends on the number of recent clicks is highly time-sensitive. We also conduct empirical analyses on real-world datasets to validate this point, with making the following two interesting observations: (1) \textit{The more popular an item is, the larger average rating value the item tends to acquire.} This observation reveals the existence of benign popularity bias --- items with higher popularity usually suggest better quality and would receive more praise. (2) \textit{From the temporal view, for a large proportion of items, the rating value exhibits negative correlation with the item popularity at that time.} This observation reveals temporal dynamic of harmful popularity bias --- conformity exerts varying negative impact on users' behaviors with time going by.

Based on the above insights, we propose a \textbf{Ti}me-aware \textbf{D}is\textbf{E}ntangled framework (\textbf{TIDE}) for tackling popularity bias. We resort to the causal graph and assume click data is generated from three different components: (1) a time-invariant module that captures the excellence of the item; (2) a temporal dynamic module that encodes the conformity effect by scrutinizing the number and time of recent clicks on the item; (3) a normal recommendation model that estimates user interest matching on the item. Such disentangled model provides opportunity to make better recommendation --- inheriting the benign components while circumventing the harmful ones. Towards this end, during the inference stage, we conduct causal intervention on the conformity module to make the prediction beneficial from the item quality and interest matching score while immune to the harmful conformity effect.

Lastly, in terms of leveraging popularity bias in recommendation, the most relevant work is the recently proposed PDA\cite{DBLP:conf/sigir/ZhangF0WSL021}.  However, we argue that directly injecting (predicted) item popularity score into prediction is insufficient for satisfactory recommendation as the harmful conformity effect is also injected. Distinct from PDA, our TIDE distills the benign popularity bias in prediction and yields significant empirical improvement.




In a nutshell, this work makes the following main contributions:
\begin{itemize}
\item To the best of our knowledge, this is the first work to study the problem of disentangling the benign popularity bias caused by item popularity from the harmful popularity bias caused by conformity in recommendation.
\item We propose a novel time-aware disentangled framework TIDE for tackling popularity bias in recommendation. TIDE performs disentangled training by leveraging temporal information while resorts to intervention to block the harmful conformity effect during inference stage.
\item Extensive experiments on three well-known benchmark datasets demonstrate the superiority of the proposed method over a range of state-of-the-arts. We will release our source code to facilitate future research.
\end{itemize}

The rest of this paper is organized as follows. We formulate the task and empirically explore popularity bias in section \ref{se:2}. We further present our proposed TIDE in section \ref{se:3}. The experimental results are presented in section \ref{se:4}. We briefly review related works in section \ref{se:5}. Finally, we conclude the paper and present some directions for future work in section \ref{se:6}.

\begin{figure*}[t!]
\centering

\subfigure[Average ratings with popularity on Douban-Movie]{
\label{fig:c1}
\includegraphics[width=0.5\textwidth]{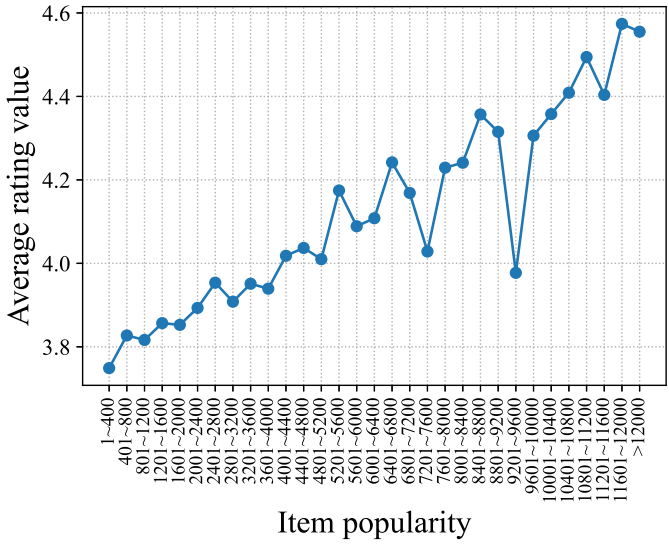}
}
\subfigure[Correlation Coefficient on various datasets]{
\label{fig:c2}
\includegraphics[width=0.45\textwidth]{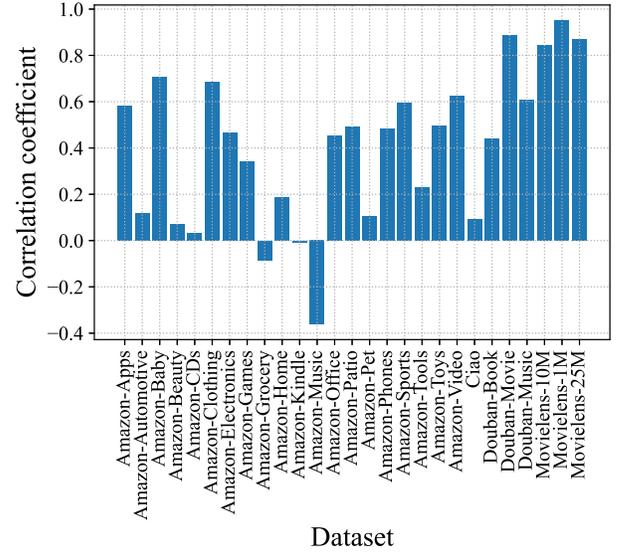}
}
 \caption{We divide items into 30 groups according to their popularity and then calculate average rating values of items in each group. The left subplot (a) shows the relation of average rating values with their popularity on Douban-Movie. The right subplot (b) presents the Correlation Coefficient between average ratings and popularity on various datasets. }
\label{fg:em1}
\end{figure*}

%

\section{Preliminaries}
\label{se:2}
In this section, we formulate the task and explore popularity bias on real-world datasets.
\subsection{Problem Definition}
\label{se:exmf}
Suppose we have a recommender system with a user set $\Set U$ and an item set $\Set I$. Let $u$ (or $i$) denotes a user (or an item) in $\Set U$ (or $\Set I$). Let $\Set D$ denote the historical user behavior data, which was sequentially collected before the time $T$ and notated as a list of triples \ie $\Set D=\{(u_k,i_k,t_k)\}_{1\le k \le |\Set D|}$, where the triple $(u_k,i_k,t_k)$ denotes the user $u_k$ has clicked the item $i_k$ at the time $t_k$. For convenience, we collect users' feedback on the specific item $i$ before time $t$ as $\Set D_i^t=\{(u_l,i_l,t_l)\in D|i_j=i, t_l<t\}$. Also, we define the popularity $P_i$ of the item $i$ as the number of observed interactions on $i$, \ie $P_i=|D_i^T|$. The task of a recommendation system can be stated as follows: learning a recommendation model from $\Set D$ so that it can capture user preference and make a high-quality recommendation.

\textit{Popularity Bias}, which denotes the uneven (usually long-tailed) distribution over the interaction frequency of items, is common in a recommender system. There are two factors resulting in popularity bias: (1) item quality, revealing the inherent excellence of items, which is benign; (2) conformity effect, describing a user tends to behave towards group norms while deviating from his own preference, which is harmful. This paper aims at disentangling the two factors such that the recommendation can benefit from the benign factor while circumvent the harmful one.
\subsection{Empirical Analyses of Popularity Bias}
\begin{figure*}[t!]
\centering
\subfigure[Coefficient distribution on the dataset Douban-Movie.]{
\label{fig:d1}
\includegraphics[width=0.45\textwidth]{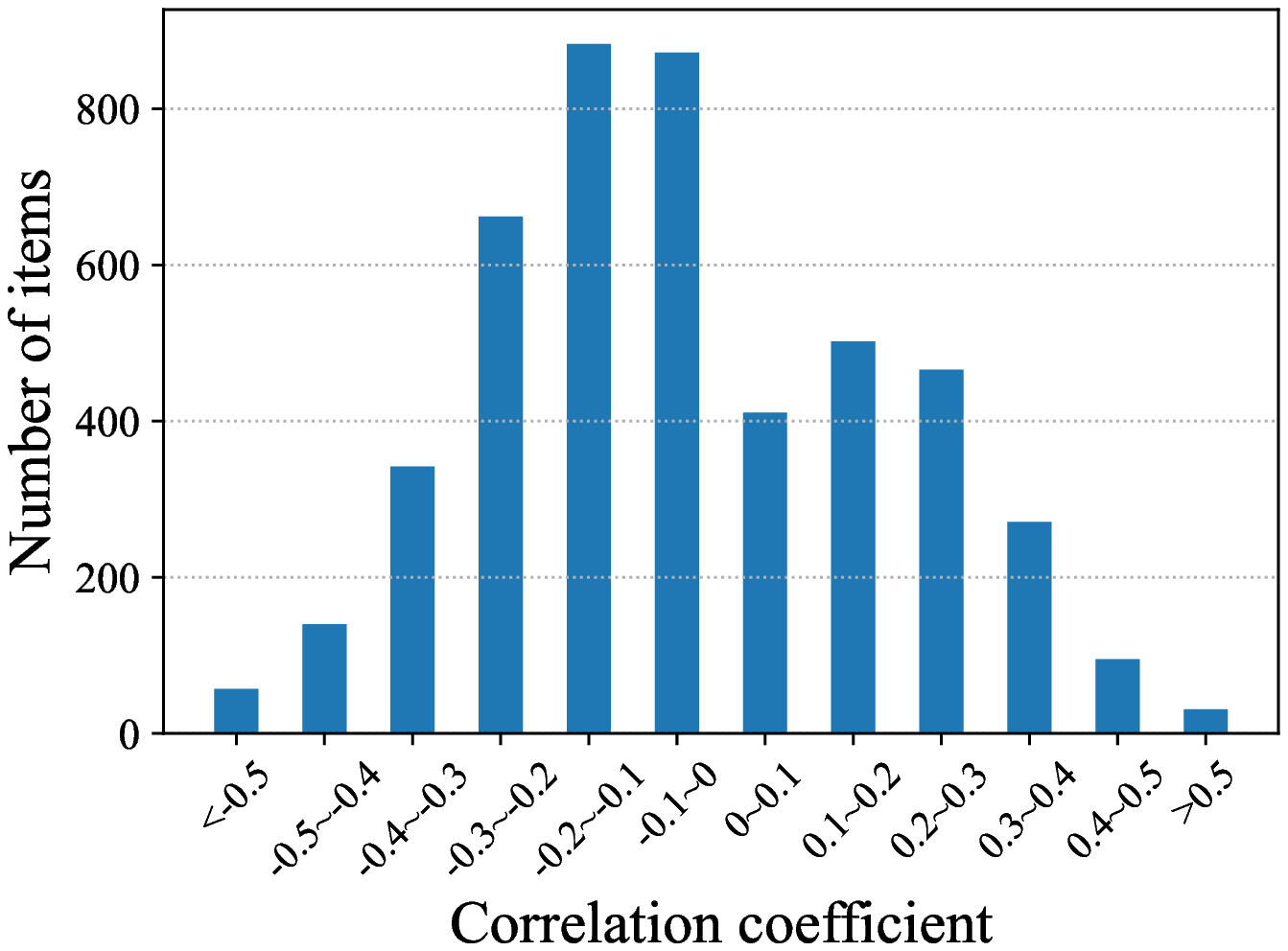}
}
\subfigure[Coefficient distribution on the dataset Amazon-CDs.]{
\label{fig:d2}
\includegraphics[width=0.45\textwidth]{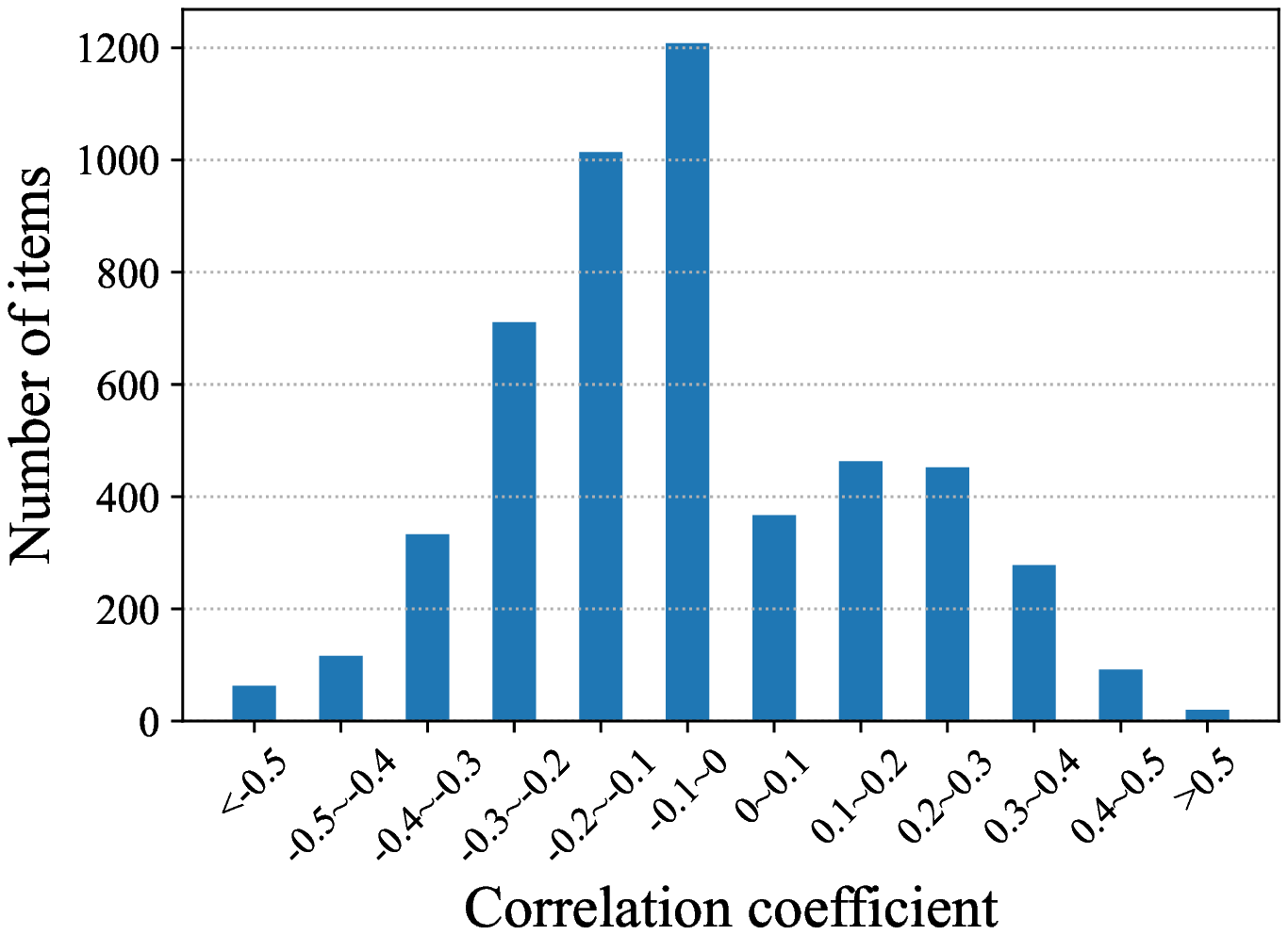}
}
\subfigure[Temporal evolving of the instant popularity for five randomly-selected items.]{
\label{fig:d3}
\includegraphics[width=0.43\textwidth]{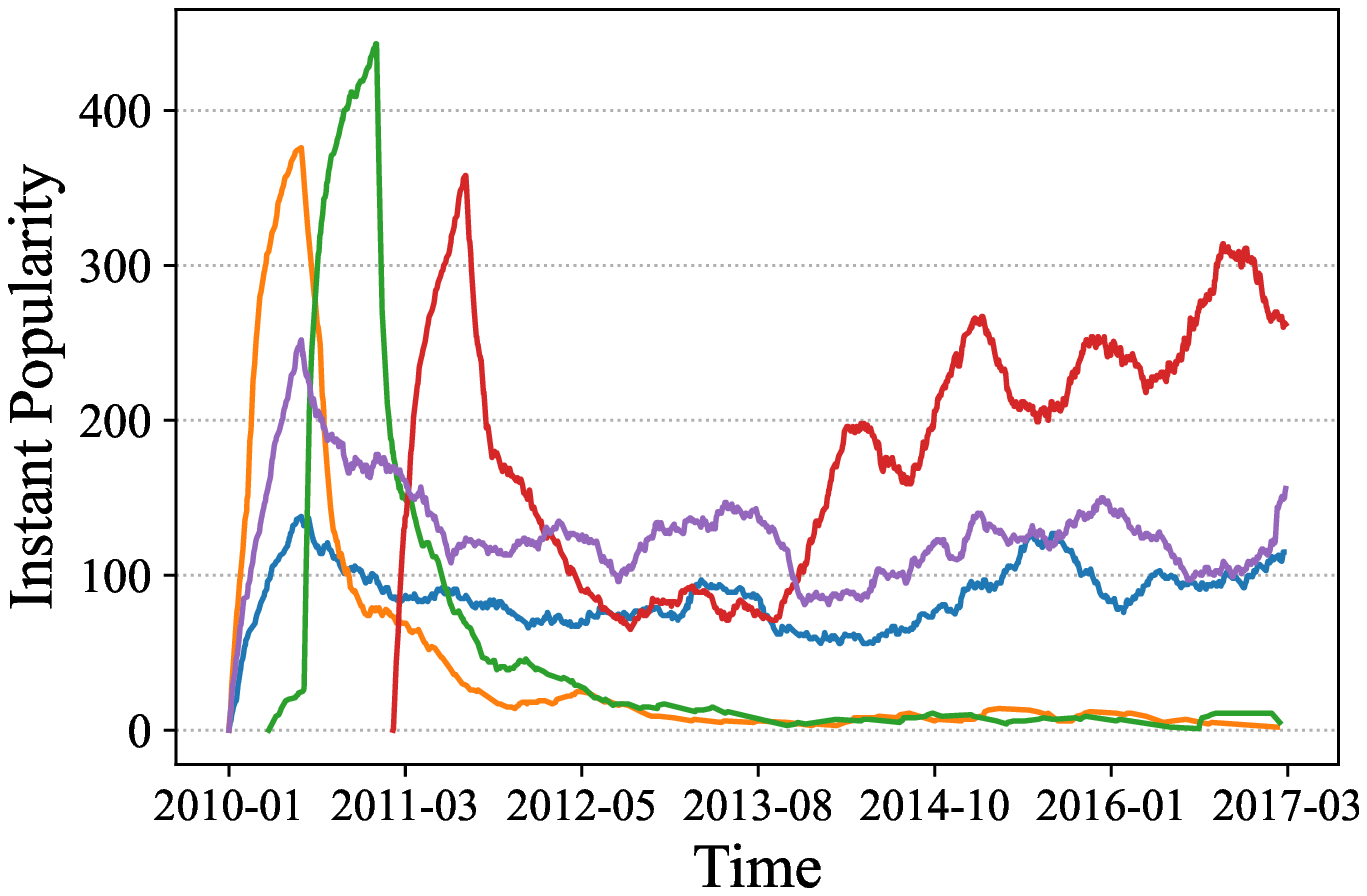}
}
\quad
\subfigure[The rating value with the instant popularity for an exemplified item. For better visualization, here we scatter the average ratings occurred within a week. Also, we plot a fitting curve for the rating value.]{
\label{fig:d4}
\includegraphics[width=0.48\textwidth]{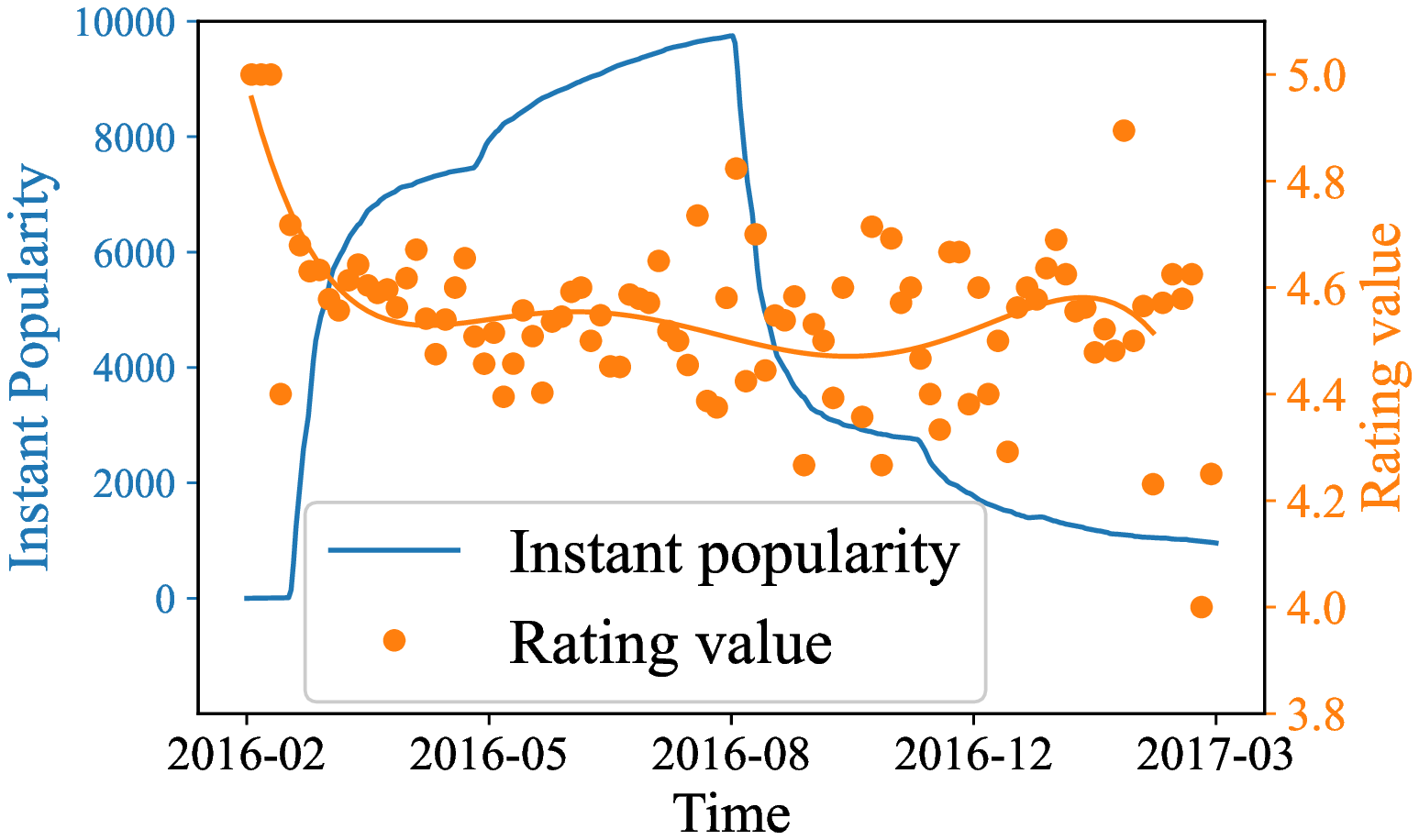}
}

 \caption{We calculate the correlation coefficient between the rating value and the instant popularity at that time for each item, where instant popularity denotes the number of clicks on the item during the past half year. The subplots (a) and (b) illustrate the distribution of the calculated coefficient over items on two typical datasets; The subplot (c) illustrates the temporal evolving of the instant popularity for five randomly-selected items on Douban-Movie; The subplot (d) visualizes the relation of the rating value with the instant popularity for an exemplified item.}
\label{fg:em2}
\end{figure*}

In this subsection, to reveal the existence of two factors and their properties, we conducted empirical analyses on real-world recommendation datasets including Amazon\footnote{\url{https://jmcauley.ucsd.edu/data/amazon/}}, Ciao\footnote{\url{https://www.cse.msu.edu/~tangjili/datasetcode/truststudy.htm}}, Douban\footnote{\url{https://github.com/DeepGraphLearning/RecommenderSystems/blob/master/socialRec/README.md\#douban-data}} and Movielens\footnote{\url{http://files.grouplens.org/datasets/movielens/}}. Besides click information, these datasets also contain users' ratings on their clicked items, which provide groud\_truth label of their preference. A larger rating value suggests a user is more satisfied with an item. Two statistical analyses have been conducted: (1) We first explore the correlation between item popularity and their ratings. We divide items into 30 groups according to their popularity (where we segment popularity interval uniformly). We then calculate the average ratings of items in each group. The result on a typical dataset Douban-Movie is presented in Figure \ref{fig:c1}. We also report the Pearson Correlation Coefficient \cite{benesty2009pearson} between the average rating and popularity in terms of groups on various datasets in Figure \ref{fig:c2}. (2) We then explore the temporal dynamic of popularity bias. For each item, we calculate the Pearson Correlation Coefficient between the rating value and the time-aware \textit{instant popularity} at that time, where \textit{instant popularity} of item $i$ and time $t$ is defined as the number of clicks on the item during the past half year (\ie $|\Set D_i^t|-|\Set D_i^{t-t_o}|$, $t_o$ denotes a period of half year). The distribution of the calculated coefficients over items on two typical datasets is presented in Figure \ref{fig:d1},\ref{fig:d2}. Here we filter out not significant results with $p>0.2$. We also visualize the temporal evolution of the instant popularity for five randomly-selected items (Figure \ref{fig:d3}), as well as an example of the relation between the rating value and the instant popularity (Figure \ref{fig:d4}).

Two important observations are concluded from these results.
\newtheorem{lemm}{\textbf{Observation}}
\begin{lemm}
The more popular an item is, the larger average rating value the item tends to have. \label{ob1}
\end{lemm}
Figure \ref{fig:c2} demonstrates item average rating values exhibit positive correlation with item popularity in a large portion of datasets. This result suggests that popularity bias is not always harmful. Some items with higher popularity can be attributed to their better intrinsic quality, which are more likely to be favored by users. Item popularity provides an imporatant signal regarding to item quality, which is profitable to boost recommendation performance. Nevertheless, item popularity can not be directly leveraged into recommendation. Popularity would also be affected by the conformity effect, deviating from the quality. It can be seen from the severe fluctuation of the curve in Figure \ref{fig:c1}. Also, popularity exhibits weakly-positive or even negative correlation with average ratings in a considerable portion of datasets. Thus, we need to disentangle the effects from the two factors such that the recommendation can benefit from such benign knowledge while circumvent the impact of the harmful one.

\begin{lemm}
From the temporal view, for a large proportion of items, the rating value exhibits negative correlation with the item temporal popularity at that time. \label{ob2}
\end{lemm}

Figure \ref{fig:d3},\ref{fig:d4} vividly demonstrates the dynamic of item instant popularity that conformity effect depends on. Besides, when the instant popularity becomes larger, when the conformity exerts larger impact on user behavior, we observe to a large extent that user's behavior deviates from his own preference. Thus we can see the negative correlation between average ratings and instant popularity (Figure \ref{fig:d1},\ref{fig:d2}). This observation reveals the temporal dynamic of harmful popularity bias and motivates us to leverage temporal information in disentanglement.

Based on above analyses, we make the following hypothesis, which lays foundation of our proposed method:
\newtheorem{lemm1}{\textbf{Hypothesis}}
\begin{lemm1}
Popularity bias is mainly caused by both conformity effect and diverse item quality. Item quality that reveals item intrinsic property is stable and static, while conformity that depends on recent clicks is highly time-sensitive.\label{hp1}
\end{lemm1}


\begin{figure}[t!]
\centering

\subfigure[Causal graph of our TIDE.]{
\label{fig:model}
\includegraphics[width=0.2\textwidth]{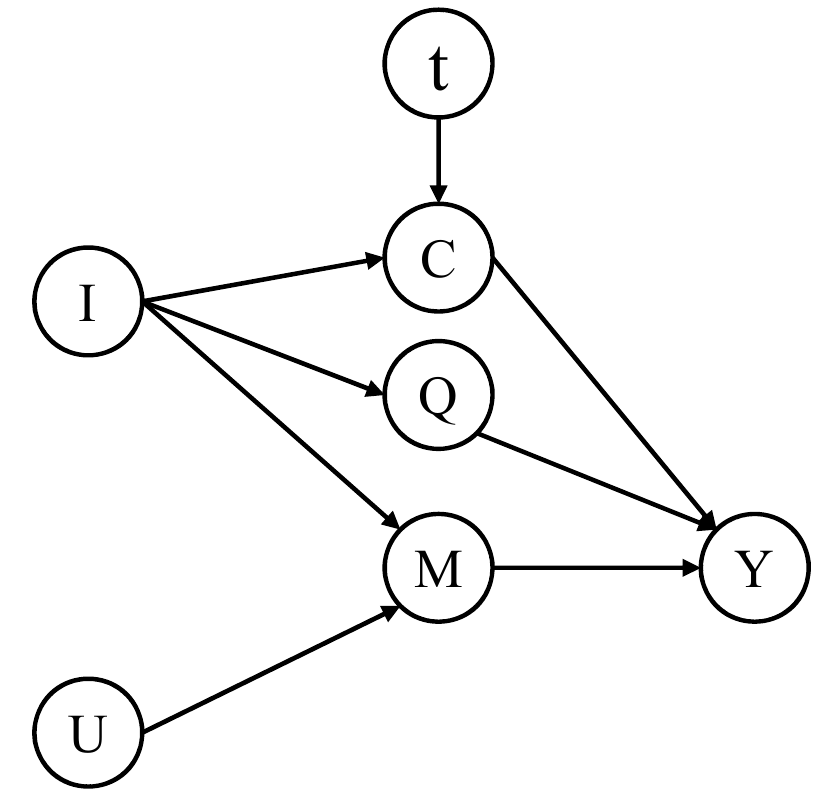}
}\quad
\subfigure[Performing intervention on $C$ during inference stage]{
\label{fig:infer}
\includegraphics[width=0.2\textwidth]{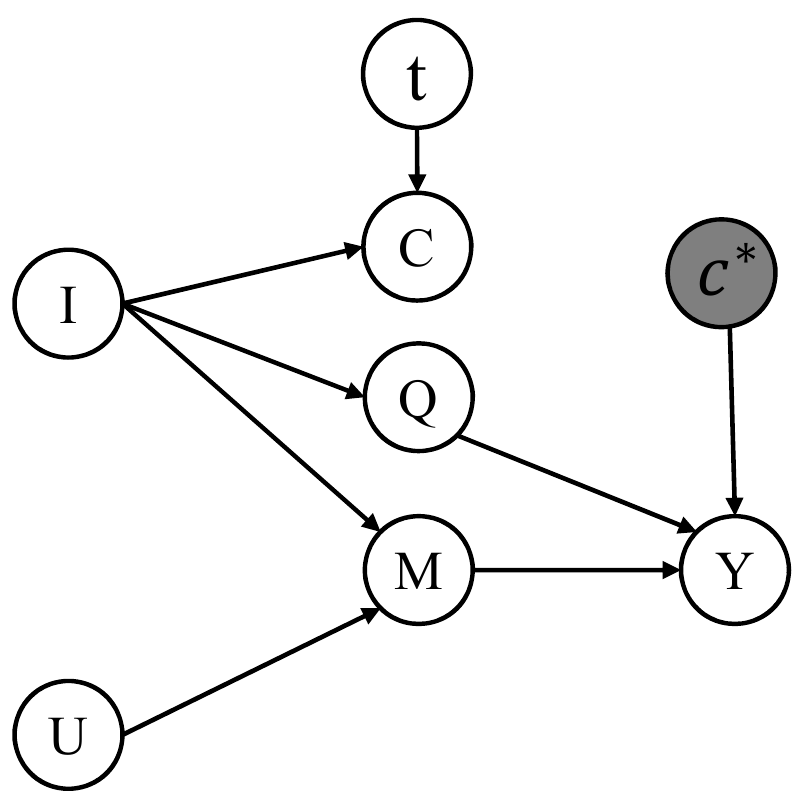}
}
\caption{The subplot (a) illustrates causal graph of TIDE while the subplot (b) illustrates how we conduct interventional inference on TIDE. I: item; U: user; C: conformity; Q: quality; M: matching score; Y: prediction.}
\label{fg:em3}
\end{figure}

\section{Time-aware Disentangled Framework}
In this section, we present our time-aware disentangled framework (TIDE) for tackling popularity bias.
\label{se:3}

\subsection{Disentangled Learning}
TIDE resorts to a causal graph as shown in Figure \ref{fig:model} and assumes an observed click is generated from the following three disentangled components:

(1) $I \to Q \to Y$ : This link denotes the effect of item quality on user behavior. An item with higher quality is more likely to be favored by a user. Here we simply use a time-irrelevant item-specific variable $q_i$ for each item $i$ to capture its inherent quality.

(2) $(I,t)\to C\to Y$: These links represent the time-aware conformity effect on user behavior. As suggested in Hypothesis \ref{hp1}, the impact of conformity not only depends on the time point $t$ of this interaction, but also on the time and the number of past interactions on the item $i$. As such, we formulate the following parameterized function $g_\beta$ to estimate the strength of conformity effect of item $i$ at time $t$:
 \begin{align}
c_i^t = {g_\beta }(t,D_i^t) = {\beta _i}\sum\limits_{({u_l},{i_l},{t_l}) \in D_i^t} {\exp ( - \frac{{|t - {t_l}|}}{\tau })}
\end{align}
where a parameter $\beta _i$ is introduced for each item $i$ to re-scale the effect, as conformity usually exhibits more severe on some items (\eg soap opera) than others (\eg science documentary). Here we simply cumulate the stimulations from past interactions while discount their contribution according to the time interval. This setting is coincident with our intuition --- the currently popular items would have larger impact on us than the one that was popular in the far past. We also introduce a temperature parameter $\tau$ controlling the sensitivity of $c_i^t$ to the time. A smaller $\tau$ would make the model focus more on recent interactions and immunize the interactions occurred long time ago.


(3) $(U,I) \to M \to Y$: these links project user and item features (\eg IDs) into their matching scores $m_{ui}=f_\theta(u,i)$. $f_\theta(u,i)$ can be implemented by various recommendation models, such as MF\cite{RendleBPR}, LightGCN \cite{he2020lightgcn}, DIN\cite{zhou2018deep}and etc.

Finally these three components are aggregated into a final prediction score for recovering the observed historical interactions:
 \begin{align}
\hat y^t_{ui}=\text{Tanh}(q_i+c_i^t) \times \text{Softplus}(m_{ui})
\end{align}
Where $\text{tanh}(q_i+c_i^t)$ can be understood as popularity bias which combines the benign effect from the item quality ($Q \to Y$) and the harmful effect from the conformity ($C \to Y$). Tanh(.) is an activation function that project the popularity bias into interval [0,1] while Softplus(.) is an activation to ensure the positivity of the matching score.

We can still apply commonly-used BPR \cite{RendleBPR} recommendation loss over the final prediction score to learn the model. Formally, the training loss is given as follow:
 \begin{align}
L = \sum\limits_{(u,i,t) \in \Set D,j \sim p_n} {-\log(\sigma (\hat y_{ui}^t - \hat y_{uj}^t))}
 \end{align}
where $\sigma$ represents the sigmoid function. We conduct negative sampling to draw a negative instance $j$ from distribution $p_n$ for training our model. As recent work \cite{DBLP:conf/sigir/ZhangF0WSL021}, here we simply use a uniform sampling strategy for fair comparison. Note that we have omitted the $L_2$ regularization terms for clarity.

\subsection{Intervention-based Inference}
As shown in Figure \ref{fig:model}, $I$ influences $Y$ through three paths: $I \to Q \to Y$ through item quality, $I \to C \to Y$ through conformity effect and $I \to M \to Y$ through user-item matching score. In order to make the recommendation benefit from the useful factors while circumvent the harmful, we perform the causal intervention to cut off the path $I \to C \to Y$ as shown in Figure\ref{fig:infer} where improper effect from the conformity has been removed. Formally, we directly intervene $c_i^t$ with a fixed value $c^*$ and make the prediction as:
 \begin{align}
\hat y_{ui}^* = {\rm{tanh}}({q_i} + {c^*}) \times {\rm{Softplus}}({m_{ui}})
  \end{align}
We set $c^*$ as 0 in experiments for the simplicity of the model.

\begin{figure}[t!]
    \centering
    \includegraphics[width=0.47\textwidth]{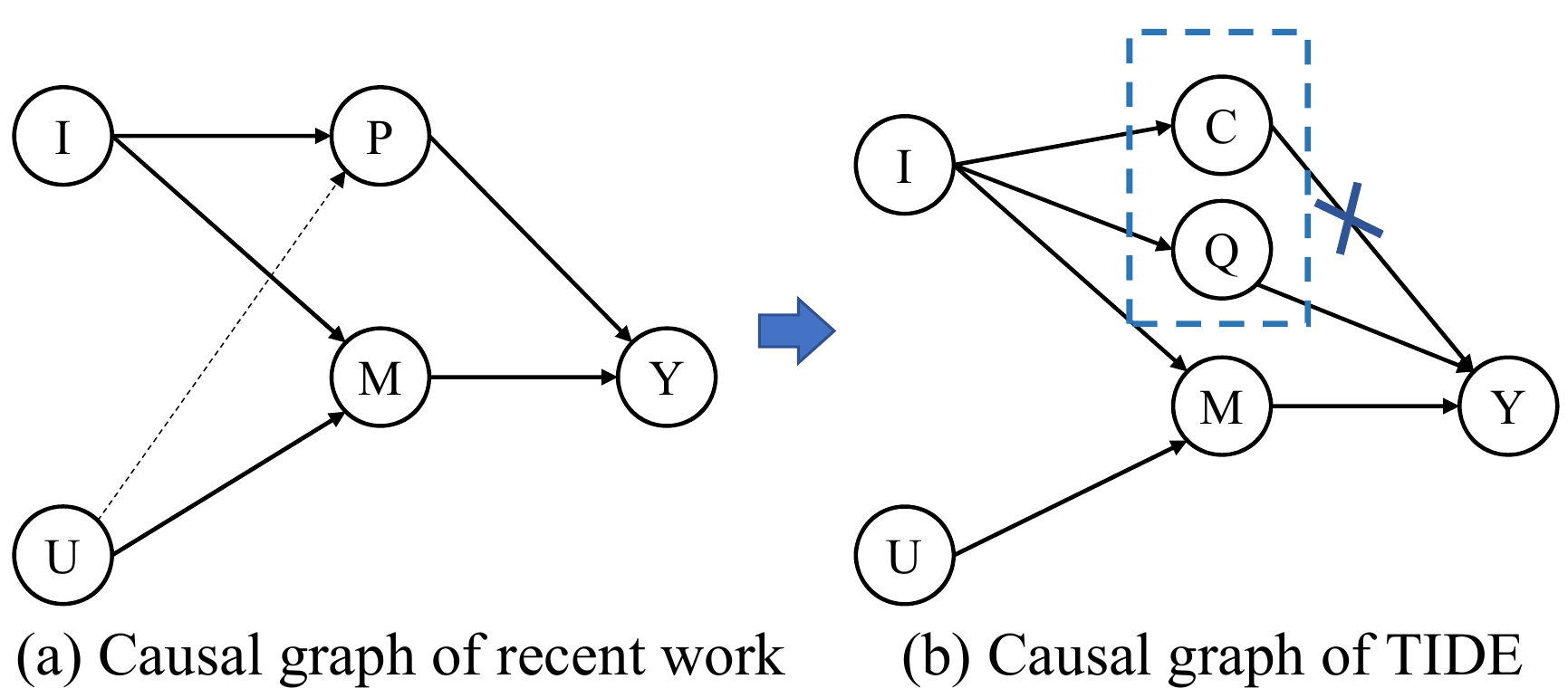}
    \caption{Causal graph comparison of recent work with our TIDE.}
    \label{fig:duibi}
\end{figure}

\subsection{Links to Recent Work}
Recent years have witnessed various debiasing strategies for popularity bias. Among which, causal inference is the most successful and representative strategy \cite{DBLP:conf/sigir/ZhangF0WSL021,zheng2021disentangling,wei2020model}. We argue that the inherent nature of this kind of methods is disentanglement --- undo the effect of the popularity bias to recover user preference on items. The cause graph of these methods can be simply summarized as Figure \ref{fig:duibi}(a). Although this graph may be different from the causal graph claimed in the original papers, Figure \ref{fig:duibi}(a) is indeed coincident with their models. For example, PDA \cite{DBLP:conf/sigir/ZhangF0WSL021} assumes a click is generated with combining item popularity score and user-item matching score, \ie ${{\hat y}_{ui}} = P_i^\gamma  \times \text{Elu}({m_{ui}})$; DICE \cite{zheng2021disentangling} makes a similar assumption except that they model sensitivity of users to item popularity (as marked by the dash line in Figure \ref{fig:duibi}(a)).

This work lies on this scheme but we further conduct disentanglement of popularity bias. As Figure \ref{fig:duibi}(b) shows, we split the path regarding to popularity bias ($I\to P \to Y$) into two paths: $I\to Q \to Y$ the benign effect from item quality and $I\to C \to Y$ the harmful effect from conformity. Besides, during the inference stage, instead of blindly removing popularity bias as \cite{zheng2021disentangling,wei2020model} (cutting $I\to P \to Y$) or leveraging complete popularity bias in prediction as \cite{DBLP:conf/sigir/ZhangF0WSL021}, we utilize partial popularity bias --- leveraging benign part (maintain path $I\to Q \to Y$) while removing harmful part (cutting path $I\to C \to Y$). In this way, our TIDE can distill useful information from popularity to prediction and yields empirical improvement over them.

\section{Experiments}
\label{se:4}
In this section, we conduct experiments to evaluate the performance of our proposed TIDE. Our experiments are intended to address the following research questions:
\begin{itemize}
    \item [\textbf{RQ1:}] Does TIDE outperform SOTA methods for popularity bias?
    \item [\textbf{RQ2:}] Is it beneficial to model both static item quality and dynamic conformity effect? Is it beneficial to remove the effect from conformity during the inference stage?
    \item [\textbf{RQ3:}] Do the learned parameters $q_i$ capture item quality?
\end{itemize}

\begin{table}[t!]
  \centering
  \caption{Statistics of the datasets.}
  \label{tb:ds}
    \begin{tabular}{c|c|c|c|c}
    \hline
    {\textbf{Dataset}} & \textbf{User \#} & \textbf{Item \#} & \textbf{Interaction \#} & {\textbf{Date}} \\
    \hline
    \hline
    Douban-Movie & 48,799 & 26,813 & 7,409,868 & 2010.1-2017.3 \\
    \hline
    Amazon-CDs & 75,258 & 64,443 & 1,097,592 & 1997.11-2014.7  \\
    \hline
    Ciao  & 5,868 & 10,724 & 143,217 &   2000.5-2011.4 \\
    \hline
    \end{tabular}
\end{table}

\subsection{Experimental Setup}
\label{experiment setup}
\textbf{Datasets.} We choose three well-known datasets Ciao, Amazon-CDs and Douban-Movie for our experiments. These datasets contain users' rating records in a chronological order. Since it is unreliable to include users with few interactions for evaluation, we conduct 5-core filtering for the datasets Ciao and Amazon-CDs, and 10-core filtering for Douban-Movie. The statistics of the datasets are described in Table \ref{tb:ds}. We follow the setting of PDA \cite{DBLP:conf/sigir/ZhangF0WSL021} and split the datasets chronologically. Specifically, we split the datasets into 10 parts according to the interaction time, and each part has the same time interval. The first nine parts are used for training, while the last part is left for validation and testing, in which the interactions of half of the users are organized as validation set while others are organized as testing set. We also transform the data into binary implicit feedback for experiments as \cite{DBLP:conf/sigir/ZhangF0WSL021,chen2019samwalker}. That is, as long as there exists a rating, the corresponding implicit feedback is assigned a value of 1, suggesting the item has been interacted (\ie clicked) by the user.

\textbf{Evaluation Methodology.} We train a model with binary training data and evaluate its performance on the following two tasks:
\begin{itemize}
\item \textit{Click prediction task:} We evaluate how a model forecasts users' future clicks. Specifically, we apply the model to sort the items that have not been interacted, and test whether the top-K items would be clicked by the user in the future (\ie in test data). For the matrics, we employ Recall@K (called CP-Rec@K in this task), Precision@K (CP-Pre@K) and Normalized Discounted Cumulative Gain@K (CP-NDCG@K) for evaluating model performance in this task.

\item \textit{Preference prediction task:} Note that click is not coincident with user preference. We further evaluate how a model retrieves relevant items that users are indeed fond of. We resort to the ground\_truth rating value, and consider the item with a high rating value (\eg 5) as positive. As we do not know user's true preference on unrated items, in this task, we just rank the rated items in the test data and evaluate whether the positive items are retrieved within Top-K positions. Specifically, precision@K (marked as PP-Pre@K) and recall@K (PP-Rec@K) are adopted in this task. Also, considering the number of rated items is usually small, we set a relatively small K (\eg K=3).
\end{itemize}

\textbf{Comparison methods.} Five type of methods are tested in our experiments:
\begin{itemize}
\item MF \cite{RendleBPR}: the basic matrix factorization model with BPR loss.
\item MF-IPS \cite{joachims2017unbiased,schnabel2016recommendations}: a classic strategy for eliminating popularity bias by re-weighting each instance according to item popularity. We refer to \cite{bottou2013counterfactual} and apply a max-capping trick on IPS value to reduce variance.
\item DICE \cite{zheng2021disentangling}: a framework that leverages cause-specific data to disentangle user preference and popularity bias into two sets of embeddings.
\item PD and PDA \cite{DBLP:conf/sigir/ZhangF0WSL021}: a state-of-the-art method that performs deconfounded training while intervenes the popularity bias during model inference. We report two versions of this work: PD that directly uses matching score for recommendation; PDA that leverages predicted item popularity score in recommendation. As PDA demonstrates superior performance over ranking-based methods \cite{abdollahpouri2019managing,zhu2021popularity}, we do not include these methods as baselines.
\item TIDE: the method proposed in this work. We mainly test two versions of TIDE: TIDE-full, combining all the effect from three components for predicting user future click, \ie we use $\hat y^T_{ui}$ for ranking; TIDE-int, which performs intervention to cut off the effect from the conformity, \ie $\hat y^*_{ui}$ is utilized.
\end{itemize}

\textbf{Implementation details.} Matrix Factorization (MF) has been selected as a backbone recommendation model for experiments, and it would be straightforward to replace it with more sophisticated models such as Factorization Machine~\cite{rendle2012factorization}, or Neural Network~\cite{he2017neural,he2020lightgcn}. We also utilize re-parameterization trick to ensure the positivity of the learned $q_i$ and $\beta_i$, \ie $q_i \leftarrow Softplus(q_i), \beta_i \leftarrow Softplus(\beta_i)$. We optimize our TIDE with Adam. Grid search is used to find the best hyper-parameters based on the performance on validation set: The search spaces of learning rate and weight decay of the parameters in MF are \{1e-4, 1e-3, 1e-2\};  Also, we set the decay of $q_i$ and $\beta_i$ as 0, and search their initialization in [-5,1] with step 1 and learning rate in \{3e-4, 1e-3, 3e-3, 1e-2\}; $\tau$ is set as 1e7, batch-size is set as 8,192. The setting of compared methods is either determined by grid search in our experiments or suggested by their original papers. All experiments are conducted on a server with 2 Intel E5-2620 CPUs, 4 NVIDIA GTX2080 GPUs and 256G RAM. We will share our source code at Github when the paper gets published.

\begin{table*}[h!]
  \centering
  \footnotesize
  \caption{Performance comparison on the click prediction task. The boldface font denotes the winner in that column.}
  \label{Performance}
  \begin{tabular}{l|c c c|c c c|c c c}
  \hline
  {\bf Datasets} & \multicolumn{3}{c|}{{\bf Douban-Movie}} & \multicolumn{3}{c|}{{\bf Amazon-CDs}} & \multicolumn{3}{c}{{\bf Ciao}} \\
  \hline
  {Metrics} & \scriptsize{CP-Rec@20} & \scriptsize{CP-Pre@20} & \scriptsize{CP-NDCG@20} &  \scriptsize{CP-Rec@20} & \scriptsize{CP-Pre@20} & \scriptsize{CP-NDCG@20} &  \scriptsize{CP-Rec@20} & \scriptsize{CP-Pre@20} & \scriptsize{CP-NDCG@20} \\
  \hline
  \hline
  MF & 0.0227 & 0.0359 & 0.0389 & 0.0130 & 0.0035 & 0.0039 & 0.0142 & 0.0090 & 0.0095 \\
  \hline
  MF-IPS & 0.0219 & 0.0349 & 0.0380 & 0.0130 & 0.0035 & 0.0039 & 0.0190 & 0.0092 & 0.0094 \\
  \hline
  DICE & 0.0219 & 0.0343 & 0.0378 & 0.0114 & 0.0028 & 0.0031 & 0.0171 & 0.0112 & 0.0123 \\
  \hline
  PD  & 0.0350 & 0.0488 & 0.0549 & 0.0147 & 0.0036 & 0.0040 & 0.0165 & 0.0103 & 0.0105 \\
  \hline
  PDA & 0.0400 & 0.0549 & 0.0618 & 0.0193 & 0.0046 & 0.0052 & 0.0247 & 0.0146 & \textbf{0.0161} \\
  \hline
  {TIDE-full} & \textbf{0.0417} & \textbf{0.0571} & \textbf{0.0648} & \textbf{0.0238} & \textbf{0.0054} & \textbf{0.0062} & \textbf{0.0258} & \textbf{0.0148} & 0.0152 \\
  \hline
  Impv & 4.20\% & 4.04\% & 4.95\% & 23.43\% & 16.45\% & 19.69\% & 4.32\% & 1.24\% & -6.13\% \\
   \hline
  \end {tabular}
\end{table*}

\begin{table}[t!]
  \centering
  \footnotesize
  \caption{Performance comparison on the preference prediction task. The boldface font denotes the winner in that column.}
  \label{Performance2}
\begin{tabular}{l|c c|c c|c c}
\hline
{\bf Datasets} & \multicolumn{2}{c|}{{\bf Douban-Movie}} & \multicolumn{2}{c|}{{\bf Amazon-CDs}} & \multicolumn{2}{c}{{\bf Ciao}} \\
\hline
{ Metrics} & \begin{tabular}[c]{@{}c@{}}PP-\\ Rec@3\end{tabular} & \begin{tabular}[c]{@{}c@{}}PP-\\ Pre@3\end{tabular} & \begin{tabular}[c]{@{}c@{}}PP-\\ Rec@3\end{tabular} & \begin{tabular}[c]{@{}c@{}}PP-\\ Pre@3\end{tabular} & \begin{tabular}[c]{@{}c@{}}PP-\\ Rec@3\end{tabular} & \begin{tabular}[c]{@{}c@{}}PP-\\ Pre@3\end{tabular} \\
\hline
\hline
    MF    & 0.1718  & 0.4422  & 0.4219  & 0.6940  & 0.2644  & 0.5487  \\
    MF-IPS & 0.1693  & 0.4345  & 0.4215  & 0.6935  & 0.2652  & 0.5641  \\
\hline
    DICE  & 0.1711  & 0.4433  & 0.4204  & 0.6925  & 0.2575  & 0.5513  \\
\hline
    PD    & 0.1652  & 0.4254  & 0.4225  & 0.6959  & \textbf{0.2727}  & 0.5846  \\
    PDA   & 0.1676  & 0.4216  & 0.4265  & 0.6996  & 0.2429  & 0.5282  \\
\hline
    TIDE-full  & 0.1582  & 0.3915  & 0.4301  & 0.7059  & 0.2602  & 0.5462  \\
    TIDE-int & \textbf{0.1770} & \textbf{0.4648} & \textbf{0.4363} & \textbf{0.7176} & 0.2699 & \textbf{0.5872} \\
\hline
\end {tabular}
\end{table}

\subsection{Performance Comparison (RQ1)}
\textbf{Performance on click prediction task.} Table \ref{Performance} presents the performance of the compared methods in the click prediction task in terms of three evaluation metrics. The boldface font denotes the winner in that column. For the sake of clarity, the row `Impv' shows the relative improvement achieved by TIDE over all the baselines. Overall, with few exception, our TIDE outperforms all compared baselines. Especially in the dataset Amazon, the improvements are quite impressive --- 23.43\%, 16.45\% and 19.69\% in terms of Precision, Recall and NDCG respectively. This result validates that TIDE can capture more precise popularity bias and thus make a more accurate prediction of users' future behavior.

\textbf{Performance on preference prediction task.} Table \ref{Performance2} presents the performance of the compared methods on preference prediction task. We have the following observations: (1) PDA which consistently outperforms PD in the click prediction task performs worse in this task. This interesting phenomenon reveals the negative impact of popularity bias. Blindly inject popularity bias without filtering its harmful ingredient would deteriorate the model performance. Similar results can be seen from the worse performance of TIDE-full than TIDE-int. (2) Overall, with few exception, our TIDE-int outperforms all compared methods in this task. This result validates the effectiveness of disentangling benign and harmful factors of popularity bias. Without disentanglement, existing methods sink into a dilemma --- they either fail to utilize the important signal of the item quality (\eg TIDE-int outperforms PD, DICE, MF-IPS), or are disturbed by the harmful conformity (\eg TIDE-int outperforms PDA and MF).

\begin{table*}[t!]
  \centering
  \caption{Characteristics of TIDE and its variants. We also report their performance on the preference prediction task.}
    \begin{tabular}{l|ccc|ccc|cc|cc|cc}
    \hline
    \multicolumn{1}{c|}{\multirow{3}{*}{Methods}} & \multicolumn{3}{c|}{\multirow{2}{*}{Training with?}} & \multicolumn{3}{c|}{\multirow{2}{*}{Inference with?}} & \multicolumn{6}{c}{Performance} \\
\cline{8-13}          & \multicolumn{3}{c|}{}  & \multicolumn{3}{c|}{} & \multicolumn{2}{c|}{Douban-Movie} & \multicolumn{2}{c|}{Amazon-CDs} & \multicolumn{2}{c}{Ciao} \\
\cline{2-13}          & \begin{tabular}[c]{@{}c@{}}Matching\\ Scores?\end{tabular}   & Quality?     & \begin{tabular}[c]{@{}c@{}}Confor-\\mity?\end{tabular}     & \begin{tabular}[c]{@{}c@{}}Matching\\ Scores?\end{tabular}   & Quality?     & \begin{tabular}[c]{@{}c@{}}Confor-\\mity?\end{tabular}     & \begin{tabular}[c]{@{}c@{}}PP-\\ Rec@3\end{tabular} & \begin{tabular}[c]{@{}c@{}}PP-\\ Pre@3\end{tabular} & \begin{tabular}[c]{@{}c@{}}PP-\\ Rec@3\end{tabular} & \begin{tabular}[c]{@{}c@{}}PP-\\ Pre@3\end{tabular} & \begin{tabular}[c]{@{}c@{}}PP-\\ Rec@3\end{tabular} & \begin{tabular}[c]{@{}c@{}}PP-\\ Pre@3\end{tabular} \\
    \hline
    \hline
    MF & ${\surd}$  & $\times$ & $\times$ & ${\surd}$ & $\times$ & $\times$ & 0.1718  & 0.4422  & 0.4219  & 0.6940  & 0.2644  & 0.5487  \\
    \hline
    TIDE-noc & ${\surd}$ & ${\surd}$ & $\times$ & ${\surd}$ & ${\surd}$ & $\times$ & 0.1720  & 0.4439  & 0.4274  & 0.7008  & 0.2536  & 0.5385  \\
    TIDE-noq & ${\surd}$ & $\times$ & ${\surd}$  & ${\surd}$ & $\times$ & ${\surd}$ & 0.1618  & 0.3992  & 0.4255  & 0.6986  & 0.2551  & 0.5462  \\
    \hline
    TIDE-e  & ${\surd}$ &  ${\surd}$ &${\surd}$ & ${\surd}$ & $\times$ &$\times$  & 0.1582  & 0.3926  & 0.4214  & 0.6952  & 0.2554  & 0.5615  \\
    TIDE-full & ${\surd}$ & ${\surd}$ &${\surd}$  & ${\surd}$ &${\surd}$ &${\surd}$ & 0.1582  & 0.3915  & 0.4301  & 0.7059  & 0.2602  & 0.5462  \\
    TIDE-int & ${\surd}$ & ${\surd}$ &${\surd}$ & ${\surd}$ &  ${\surd}$ & $\times$ & {\bf0.1770}  & {\bf0.4648}  & {\bf0.4363}  & {\bf0.7176}  & {\bf0.2699}  & {\bf0.5872}  \\
    \hline
    \end{tabular}
  \label{tab:ab}
\end{table*}

\subsection{Ablation Study (RQ2)}
We conduct ablation study to explore whether it is essential to model both factors and whether it is essential to perform interventional inference. We compare our TIDE-full and TIDE-int with the following special cases: (1) TIDE-noq and TIDE-noc: where item quality or conformity effect is removed in both training and inference stage; (2) TIDE-e: which is trained as same as TIDE-int but uses matching score for recommendation. The characteristics and performance on preference prediction task are presented in Table \ref{tab:ab}.

\textbf{Effectiveness of modeling both factors.} We observe that the method with modeling two factors (TIDE-int) consistently outperforms the cases just considering one aspect (TIDE-noq and TIDE-noc). This result is coincident with our intuition --- modeling both factors is beneficial for capturing popularity bias as well as for distilling useful knowledge from it.

\textbf{Effectiveness of interventional inference.} From Table \ref{tab:ab}, we observe TIDE-int is consistently superior over TIDE-e and TIDE-full. This result demonstrates the mix nature of popularity bias --- containing both benign and harmful signals. The model that roughly maintains or removes both of them would result in undesirable performance.

\begin{figure}[t!]
    \centering
    \includegraphics[width=0.45\textwidth]{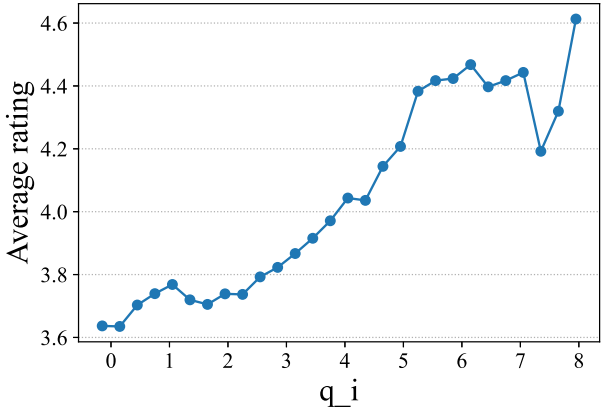}
    \caption{We divide items into 30 groups according to their learned $q_i$ and then calculate average rating values of items in each group. This figure visualizes the relation of the average rating value with learned $q_i$ on Douban-Movie. }
    \label{fig:q-ar}
\end{figure}

\subsection{Exploratory Analysis (RQ3)}
To answer the question RQ3, we now explore the learned $q_i$ from two perspective to provide insights into how TIDE captures item quality.

\textbf{Distribution of learned $q_i$.} Figure \ref{fig:q-ar} visualizes the distribution of the learned $q_i$ with their average rating value (simply marked as $AR_i$) on a typical dataset Douban-Movie. We can observe the positive correlation between them. Also, comparing with Figure \ref{fig:c1}, the curve in Figure \ref{fig:q-ar} is more stable and exhibits less fluctuation.

\begin{table}[t!]
  \centering
  \caption{Ranking Correlation Coefficient between the lists ranked by average rating $AR_i$ and by the learned parameter $q_i$, as well as with between the lists ranked by average rating $AR_i$ and by popularity $P_i$.}
    \begin{tabular}{l|c|c|c}
    \hline
          & Douban-Movie & Amazon-CDs & Ciao \\
    \hline
    $AR_i$ with $q_i$ & 0.1268  & 0.1038  & 0.0808  \\
    \hline
    $AR_i$ with $P_i$ & 0.0384  & -0.0648  & -0.0336  \\
    \hline
    \end{tabular}
  \label{tb:rcc}
\end{table}

\textbf{Ranking correlation comparison.} We further validate the stronger correlation of the average rating value with $q_i$ than with popularity $P_i$. We calculate Kendall Tau Ranking Correlation Coefficient (RCC) \cite{abdullah1990robust} between the item lists ranked by them. RCC essentially measures the probability of two items being in the same order in the two ranked lists, and would be more robust and rational than Pearson Correlation Coefficient (PCC) especially for a recommendation task. The result is presented in Table \ref{tb:rcc}. We observe RCC between $q_i$ and $AR_i$ is larger than RCC between $P_i$ and $AR_i$. Besides, to our surprise, we observe the absolute values of both metrics are relatively small. More seriously, RCC between $P_i$ and $AR_i$ is negative on the datasets Amazon-CDs and Ciao. This result validates the challenging of tackling popularity bias. There exists a gap between the value and ranking --- positive correlation in terms of value may not result in positive correlation in ranking. Although popularity exhibits positive correlation with $AR_i$ in PCC, its ranking result is easily distorted by other factors in popularity and deviates from reflecting positive correlation. TIDE filters conformity effect from popularity bias and relatively capture more stable and precise knowledge of item quality.

\begin{figure}[t!]
    \centering
    \includegraphics[width=0.48\textwidth]{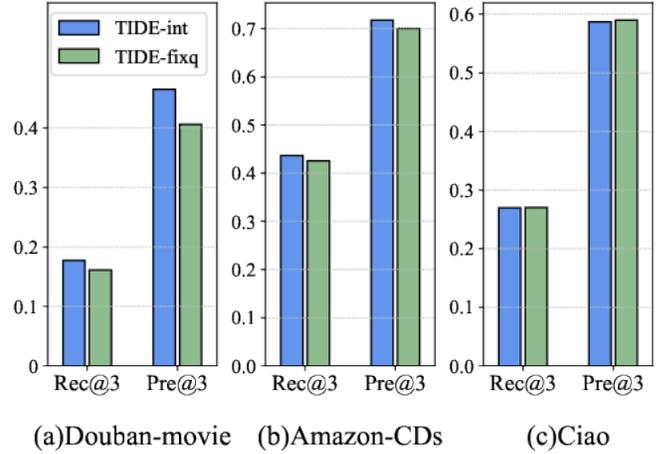}
    \caption{Comparison of TIDE-int with its special case TIDE-fixq, where all $q_i$ are constrained with the same value. }
    \label{fig:fixq}
\end{figure}

\textbf{Effectiveness of learning diverse $q_i$.} To validate the necessary of learning diverse $q_i$, we compare TIDE-int with its special case TIDE-fixq, where $q_i$ for all items are fixed as a same value. The results are presented in Figure \ref{fig:fixq}. Generally speaking, TIDE-int performs better than TIDE-fixq. Especially on two large datasets Douban-Movie and Amazon-CDs, TIDE-int consistently outperforms TIDE-fixq with a certain margin. This result demonstrates that our model indeed learns some useful information, which is beneficial for recommendation. Also, we observe quite similar performance of TIDE and TIDE-fixq on Ciao. This phenomenon can be explained as follow: It seems that popularity bias on Ciao is mainly caused by conformity. conformity is quite severe and overwhelms the contribution of item quality. it can seen from the weakly-positive correlation of the $q_i$ and $AR_i$ presented in Figure \ref{fig:c2}.

\section{Related work}
\label{se:5}
In this section, we review the most related works from the following two perspectives.

\textbf{Popularity Bias in recommendation.} Popularity bias depicting uneven (usually long-tailed) distribution over the interaction frequency of items, is common in a recommender system. The negative impact of popularity bias has been studied in a large number of recent literatures. For example, some work \cite{zheng2021disentangling,canamares2018should} argues that such skewed distribution may be caused by user conformity, deviating from reflecting users' true preference. As such, recommendation models trained on such biased data would give skewed prediction. Worse still, recommendation model not only inherit the bias, but also amplifies bias, making the popular items dominate the top recommendations \cite{anderson2006long, abdollahpouri2019managing, mansoury2020feedback,brynjolfsson2006niches, celma2008hits, park2008long}. This phenomenon has been empirically verified by Abdollahpouri ~\cite{abdollahpouri2020multi}. They find popular items are recommended to a much greater degree than even what their initial popularity warrants. It would decrease serendipity \cite{ge2020understanding,lu2012serendipitous,chaney2018algorithmic} and fairness \cite{abdollahpouri2020connection,abdollahpouri2019impact,abdollahpouri2019unfairness,abdollahpouri2019managing} of recommendation results, hurting user experience and causing customer churn.

Recent work on tackling popularity bias can be mainly classified into four types: (1) Inverse propensity scoring (IPS) \cite{gruson2019offline, schnabel2016recommendations} is a classic strategy that directly adjust the data distribution with re-weighting each instance according to item popularity. (2) Ranking adjustment is another type of method \cite{abdollahpouri2019managing,zhu2021popularity} that directly re-rank the recommendation list to improve the recommendation opportunity of unpopular items. Although simple and straightforward, this type of methods rely on human heuristical design and usually sacrifices recommendation accuracy. (3) Regularization has been introduced by some researchers to push the model towards balanced recommendation \cite{chen2020esam,bonner2018causal,kamishima2014correcting,oh2011novel}. For example, Chen \etal \cite{chen2020esam} leverage regularization to transfer the knowledge from these well-trained popular items to the long-tail items; Bonner \etal \cite{bonner2018causal} leverage regularization to distill knowledge from the uniform data for addressing popularity bias. (4) Causal inference has been leveraged for addressing popularity bias. These kinds of methods mainly assume the generative process of the data with causal graphs and then disentangle the popularity bias from the user preference accordingly\cite{DBLP:conf/sigir/ZhangF0WSL021,zheng2021disentangling,wei2020model}.

However, most of existing methods focus on eliminating popularity bias. In fact, popularity bias is not always evil. It may not only result from the users' conformity to the group, but also from item quality. It would be valuable to leverage such important signal in boosting recommendation performance. To the best of our knowledge, only one work 
\cite{DBLP:conf/sigir/ZhangF0WSL021} considers to leverage popularity bias into recommendation. However, they directly injecting (predicted) item popularity score into prediction, which is insufficient for satisfactory recommendation as the harmful conformity effect is also injected. Different from these work, we consider the double-edged nature of popularity bias. We aim at disentangling the benign popularity bias from the harmful one, such that the recommendation can benefit from the merit while circumvent the harmful.

\textbf{Biases in recommendation.} Besides popularity bias, recent work has studied other type of biases in recommendation including: Selection bias, which happens as users are free to choose which items to rate, so that the observed ratings are not a representative sample of all ratings ~\cite{chen2020bias,marlin2009collaborative, jiawei2018social,hernandez2014probabilistic}; Exposure bias, which happens in implicit feedback data as users are only exposed to a part of specific items \cite{chen2020bias,abdollahpouri2020multi,liang2016modeling,chen2018modeling}; Position bias, which happens as users tend to interact with items in higher position of the recommendation list \cite{chen2020bias,ovaisi2020correcting}; Unfairness \cite{ekstrand2021exploring,stoica2018algorithmic}, which denotes the system systematically and unfairly discriminates against certain individuals or groups of individuals in favor others. Generally, there are substantial work on addressing these biases issues. We encourage the readers refer to the survey \cite{chen2020bias} for more details.

\textbf{Disentanglement in recommendation.} In terms of disentanglement, existing efforts can be classified into two lines. The first type of methods is designed for debiasing. As discussed above, this type of methods aim at disentangling user true preference from the various data biases \cite{zheng2021disentangling,wei2020model}. Another type of methods lie in disentangled representation learning. This kind of methods aim at learning a finer-granularity representation of users and items, which is beneficial for robust and explainable recommendation. For example, Ma \etal \cite{ma2019learning} leverage Variational Auto-Encoder \cite{kingma2013auto} to disentangle high-level concepts associated with user intentions as well as low-level factors (\eg size or color of a shirt). Similarly, Wang \etal \cite{wang2020disentangled} learn disentangled user representation with the merits of the interaction graph.

\section{Conclusion}
\label{se:6}

 This paper studies an important but unexplored problem --- how to disentangle the benign popularity bias caused by item quality from the harmful popularity bias caused by conformity. We first conduct empirical analyses on real-world datasets and observe quite different patterns of these two factors along time: item quality revealing item inherent property is stable and static while conformity that depends on item recent clicks is highly time-sensitive. We then propose a novel time-aware disentangled framework ({TIDE}), where a click is generated from three components namely the static item quality, the dynamic conformity effect, as well as the user-item matching score. We further provide an interventional inference strategy such that the recommendation can benefit from the benign popularity bias while circumvent the harmful one. Extensive experiments on three real-world datasets demonstrated the effectiveness of the proposed disentangled models as well as its interventional inference strategy.

One interesting direction for future work is to explore a more sophisticate conformity model $g_\beta$, which could capture more complex patterns and potentially achieve better performance than simple sum-exponential structure. Besides, this work demonstrates popularity bias is double-edged. We believe other biases may also have this nature. It will be valuable to transfer the experience of this work to tackling other biases and to explore their benign and harmful effect on recommendation.

\appendices

\ifCLASSOPTIONcompsoc
  \section*{Acknowledgments}
\else
  \section*{Acknowledgment}
\fi
This work is supported by the National Natural Science Foundation of China (U19A2079),  National Key Research and Development Program of China (2020AAA0106000), USTC Research Funds of the Double First-Class Initiative (WK2100000019), and the Meituan Inc. through Research Cooperation Project.

\ifCLASSOPTIONcaptionsoff
  \newpage
\fi

\bibliographystyle{IEEEtran}
\bibliography{sigproc}

\begin{IEEEbiography}[{\includegraphics[width=1in,height=1.25in,clip]{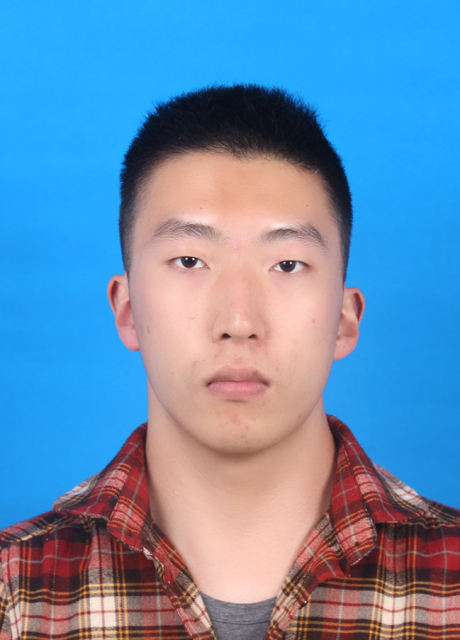}}]{Zihao Zhao}
is currently a Master Degree student with the School of Information Science and Technology, University of Science and Technology of China. He received his bachelor degree from University of Science and Technology of China in 2020. He has been awarded 2019 National Encouragement scholarship. His research interest includes recommendation system and graph representation learning.
\end{IEEEbiography}

\begin{IEEEbiography}[{\includegraphics[width=1in,height=1.25in,clip]{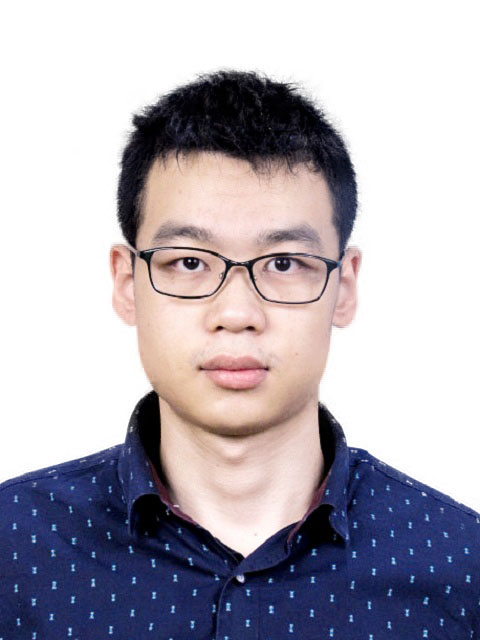}}]{Jiawei Chen}
is a Postdoc Research Fellow in School of Information Science and Technology, University of Science and Technology of China. He received Ph.D. in Computer Science from Zhejiang University in 2020. His research interests include information retrieval, data mining, and causal inference. He has published over ten academic papers on international conferences and journals such as WWW, AAAI, SIGIR, CIKM, ICDE and TOIS. Moreover, he has served as the PC/SPC member for top-tier conferences including SIGIR, WWW, WSDM, ACMMM, AAAI, IJCAI and the invited reviewer for prestigious journals such as TNNLS, TKDE, TOIS.
\end{IEEEbiography}

\begin{IEEEbiography}[{\includegraphics[width=1in,height=1.25in,clip]{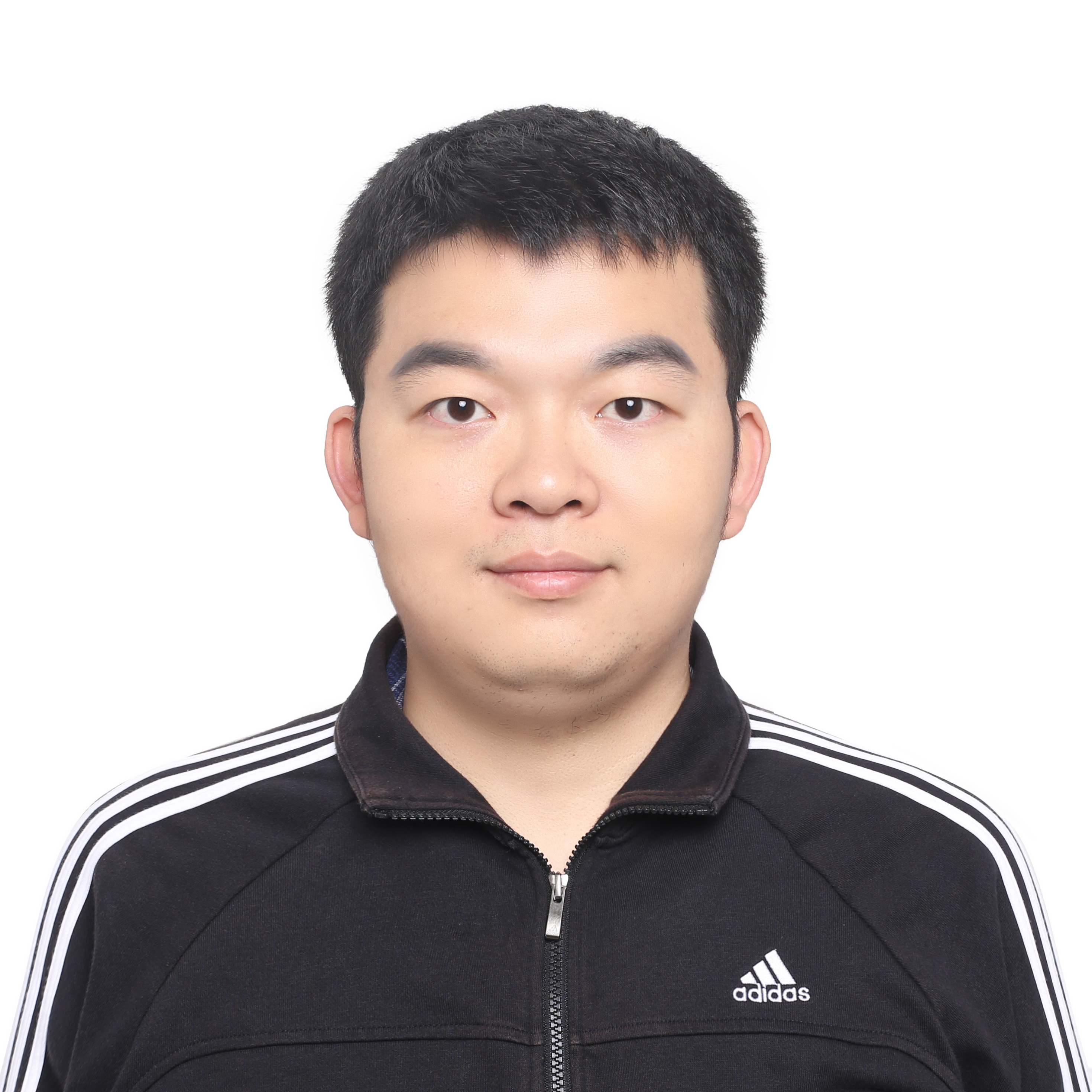}}]{Sheng Zhou}
is currently working as an assisant professor in College of Software and Engineering, Zhejiang University. He received the Ph.D degree with the College of Computer Science and Technology, Zhejiang University, No. 38 Zheda Road, Hangzhou, Zhejiang, China. He is working with Zhejiang Provincial Key Laboratory of Service Robot, College of Computer Science, Zhejiang University. His current research interests include Data mining, Graph Neural Networks and Knowledge distillation.
\end{IEEEbiography}

\begin{IEEEbiography}[{\includegraphics[width=1in,height=1.25in,clip]{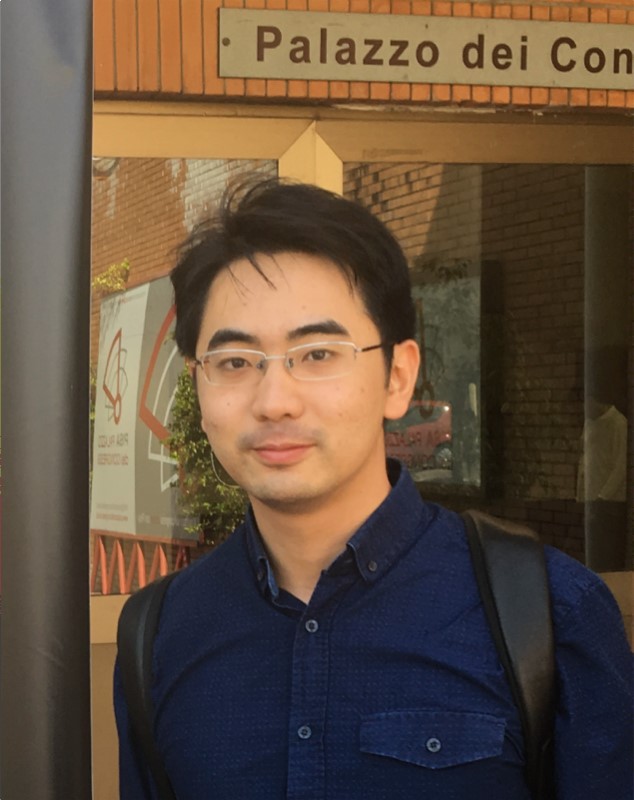}}]{Xiangnan He}
 is a professor at the University of Science and Technology of China (USTC). He received his Ph.D. in Computer Science from the National University of Singapore (NUS). His research interests span information retrieval, data mining, and multi-media analytics. He has over 100 publications that appeared in top conferences such as SIGIR, WWW, and KDD, and journals including TKDE, TOIS, and TNNLS. His work has received the Best Paper Award Honorable Mention in WWW 2018 and ACM SIGIR 2016. He serves as the associate editor for ACM Transactions on Information Systems (TOIS), Frontiers in Big Data, AI Open etc. Moreover, he has served as the PC chair of CCIS 2019 and SPC/PC member for several top conferences including SIGIR, WWW, KDD, MM, WSDM, ICML etc., and the regular reviewer for journals including TKDE, TOIS, etc.
\end{IEEEbiography}

\begin{IEEEbiography}[{\includegraphics[width=1in,height=1.25in,clip]{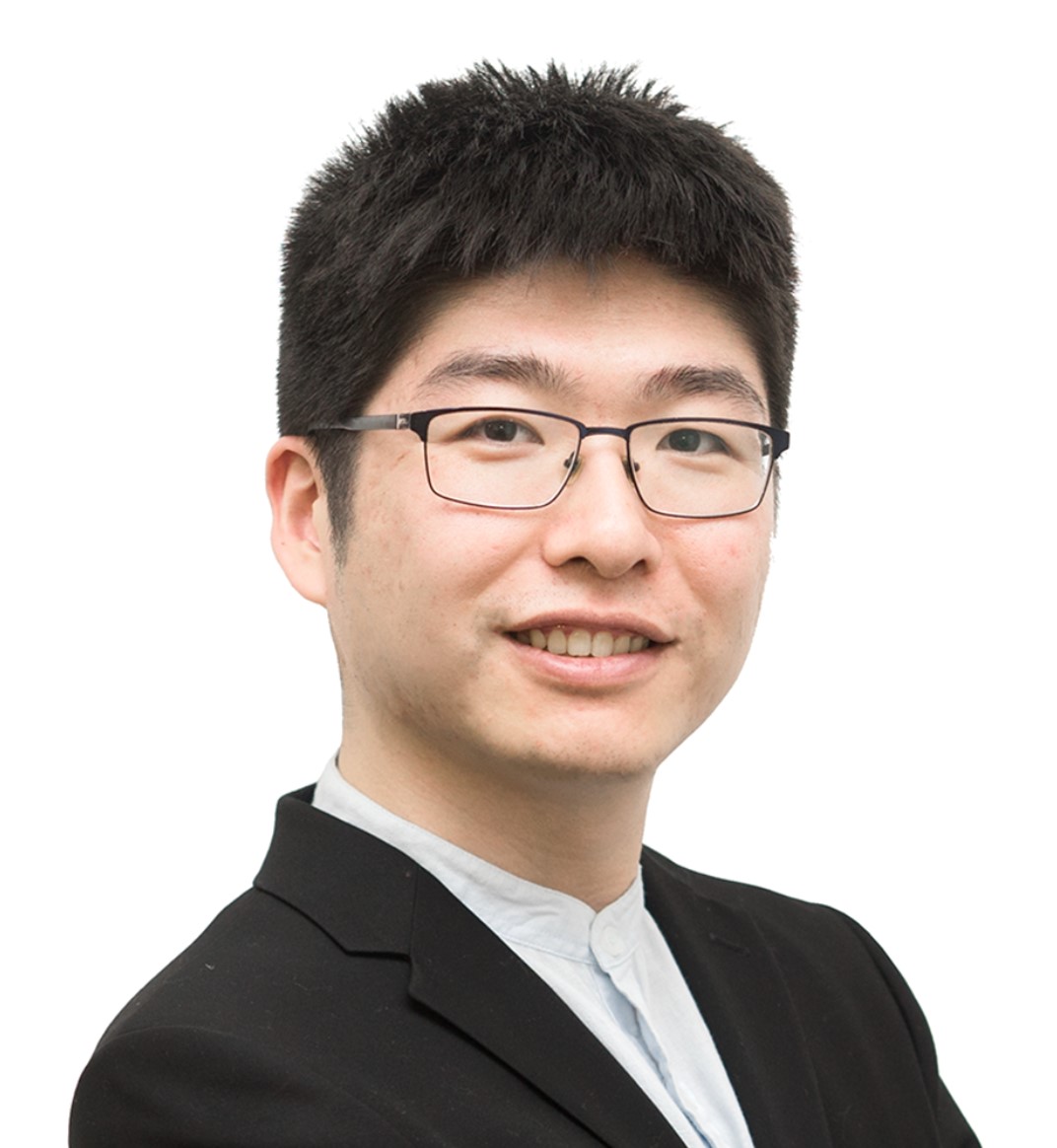}}]{Xuezhi Cao}
is a senior researcher in Meituan’s NLP team. He obtained his Ph.D. degree from Shanghai Jiao Tong University in 2018. His research interests include knowledge graph, recommender system, and social network. He has over 15 publications in top conferences including SIGIR, WWW, AAAI, etc.
\end{IEEEbiography}

\begin{IEEEbiography}[{\includegraphics[width=1in,height=1.25in,clip]{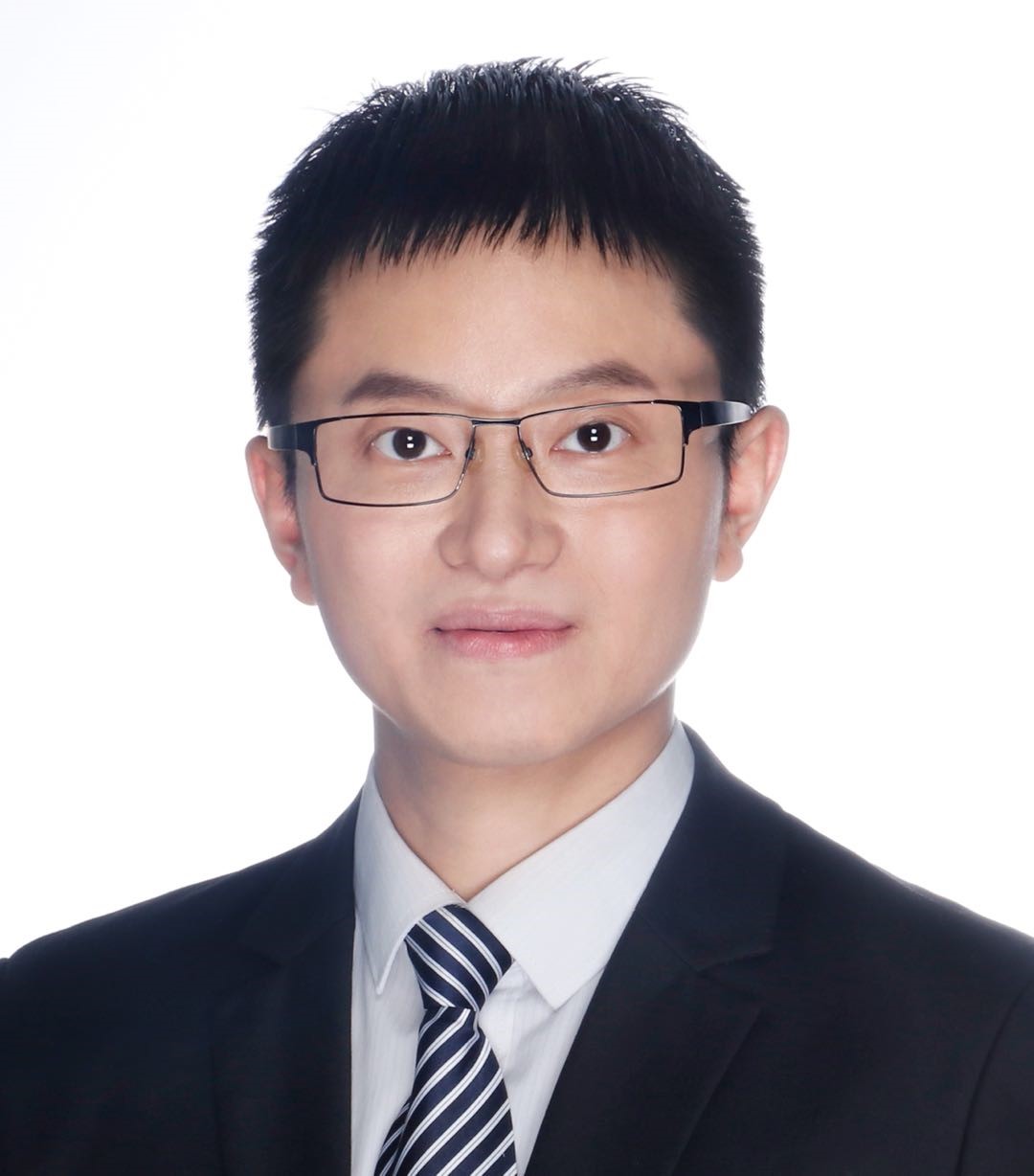}}]{Fuzheng Zhang}
is now a technical leader in Meituan. He is leading a team focus on AI technologies such as knowledge graph, NLP, and information retrieval. Before this, Dr. Zhang was a researcher in Microsoft Research Asia. He obtained his Ph.D. degree in computer science, and is supervised jointly by University of Science and Technology of China and Microsoft Research Asia. He has published over 50 top-tier international conference papers and journal articles including KDD, WWW, AAAI, and IJCAI. He has received the best paper award in ICDM2013 and CIKM2020. He has long served as the reviewers on top-tier international conferences and journals, such as KDD, WWW, and TKDE.
\end{IEEEbiography}

\begin{IEEEbiography}[{\includegraphics[width=1in,height=1.25in,clip]{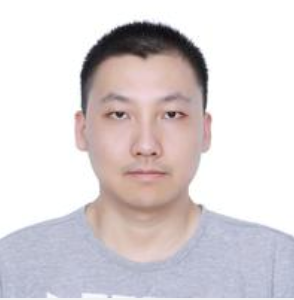}}]{Wei Wu}
is now a technical leader in Meituan. He is leading a team focus on AI technologies such as knowledge graph, NLP, and information retrieval. Before this, Dr. Zhang was a researcher in Microsoft Research Asia. He obtained his Ph.D. degree in computer science, and is supervised jointly by University of Science and Technology of China and Microsoft Research Asia. He has published over 50 top-tier international conference papers and journal articles including KDD, WWW, AAAI, and IJCAI. He has received the best paper award in ICDM2013 and CIKM2020. He has long served as the reviewers on top-tier international conferences and journals, such as KDD, WWW, and TKDE.
\end{IEEEbiography}

%
%
%
%
%
%
%
%





\end{document}